\documentclass{Interspeech}
\usepackage{cite}
\usepackage{amsmath,amssymb,amsfonts}
\usepackage{algorithmic}
\usepackage{graphicx}
\usepackage{textcomp}
\usepackage{xcolor}
\usepackage{multicol,multirow,booktabs,hyperref}

\interspeechcameraready

\title{SynHate: Detecting Hate Speech in Synthetic Deepfake Audio}

\author[affiliation={1}]{Rishabh}{Ranjan*}
\author[affiliation={2}]{Kishan}{Pipariya*}
\author[affiliation={1}]{Mayank}{Vatsa}
\author[affiliation={1}]{Richa}{Singh}

\affiliation{}{Indian Institute of Technology Jodhpur}{India}
\affiliation{}{Pandit Deendayal Energy University}{India}
\email{\{ranjan.4, mvatsa, richa@iitj.ac.in\}, kishan.pce21@sot.pdpu.ac.in}

\usepackage{comment}

\begin{document}

\maketitle

\begin{abstract}
The rise of deepfake audio and hate speech, powered by advanced text-to-speech, threatens online safety. We present \textbf{SynHate}, the first multilingual dataset for detecting hate speech in synthetic audio, spanning 37 languages. SynHate uses a novel four-class scheme: \textit{Real-normal, Real-hate, Fake-normal, and Fake-hate}. Built from MuTox and ADIMA datasets, it captures diverse hate speech patterns globally and in India. We evaluate five leading self-supervised models (Whisper-small/medium, XLS-R, AST, mHuBERT), finding notable performance differences by language, with Whisper-small performing best overall. Cross-dataset generalization remains a challenge. By releasing SynHate and baseline code, we aim to advance robust, culturally sensitive, and multilingual solutions against synthetic hate speech. The dataset is available at \url{https://www.iab-rubric.org/resources}.
\end{abstract}
\def\thefootnote{*}\footnotetext{These authors contributed equally to this work}\def\thefootnote{\arabic{footnote}}

\section{Introduction}

In the digital age, social media has become an integral part of global communication, connecting over half of the world's population. This unprecedented connectivity, however, has also fostered the spread of hate speech and malicious content. Recent surveys indicate that over one-third of social media users have encountered hate speech\footnote{https://tinyurl.com/growgate}, highlighting a pressing societal challenge. The shift from predominantly text-based interactions to multimodal content, including images, audio, and video, has further complicated the online discourse. Among these, synthetic audio produced by advanced AI technologies poses a particularly insidious threat, as it can generate highly convincing hate speech that is difficult to distinguish from genuine content.

Recent advancements in generative AI, particularly in speech synthesis, have introduced a new dimension to this challenge. State-of-the-art text-to-speech (TTS) models, such as WaveNet\cite{oord2016wavenet}, Tacotron\cite{wang2017tacotron}, and emerging diffusion-based architectures have significantly improved the quality and naturalness of synthetic speech. While these technological breakthroughs benefit numerous applications, they also enable the generation of fake audio capable of propagating hate speech. For instance, a recent incident involving a generated audio clip of Bollywood actor Ranveer Singh\footnote{https://tinyurl.com/toihate}, falsely endorsing a political party during the Indian elections, highlights the potential for misuse. Such synthetic content not only undermines trust in digital media but also risks inciting violence, manipulating public opinion, and deepening social divisions.

Audio spoofing detection focuses on distinguishing authentic speech from synthetically generated audio, with early methods relying primarily on acoustic feature extraction techniques \cite{deitcap}. Recent progress in the field has shifted towards leveraging raw audio waveforms and advanced neural architectures to improve detection robustness \cite{DBLP:conf/icb/RanjanVS22,DBLP:conf/ijcai/RanjanV023, context}. Benchmark datasets such as ASVSpoof \cite{liu2023asvspoof} and ADD \cite{yi2022add} have played a significant role in advancing the field, yet their primary emphasis remains on the signal-level characteristics of audio, without addressing the semantic content crucial for applications like hate speech detection. Furthermore, these datasets and their associated models often struggle to generalize across diverse languages and accents, limiting their effectiveness in multilingual or cross-lingual scenarios~\cite{fluent}.

Addressing these emerging threats necessitates the rapid development of robust detection mechanisms. However, progress is constrained by the absence of integrated datasets that concurrently handle hate speech detection and audio spoofing across multiple languages. Although several datasets have contributed to individual aspects of this challenge - DeToxy \cite{ghosh2021detoxy} for English hate speech; ADIMA \cite{gupta2022adima} covering 10 Indic languages; and MuTox \cite{costa2024mutox} spanning 30 languages - their focus on authentic speech limits their utility against AI-generated content. Likewise, audio spoofing detection datasets like ASVSpoof \cite{liu2023asvspoof} and ADD \cite{yi2022add} do not address the semantic content essential for hate speech classification.

\begin{figure}[t!]
    \centering
    \includegraphics[width = 0.9\columnwidth]{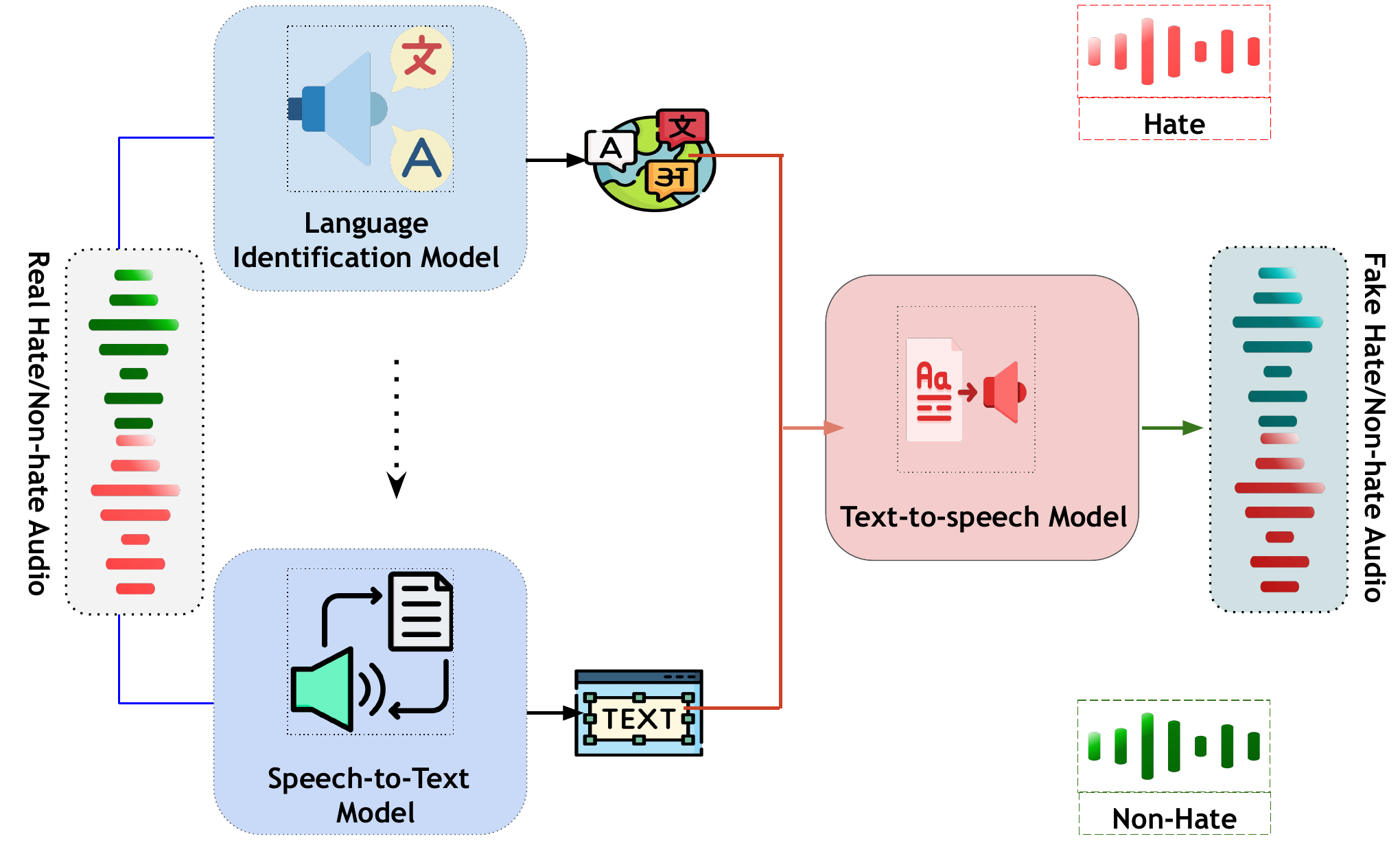}
    \caption{Pipeline for generating the proposed dataset SynHate. We use the Massive-Multilingual Speech model for language identification, speech-to-text, and text-to-speech tasks.}
    \label{fig:proposed_dataset_pipeline}
\end{figure}

To bridge this critical gap, we introduce the SynHate Dataset - the first resource designed for the simultaneous detection of hate speech and spoofed audio across multiple languages. By unifying these two tasks, SynHate tackles the complex challenge of identifying AI-generated hate speech. Our dataset employs a novel four-class categorization system: real-normal, real-hate, fake-normal, and fake-hate, which facilitates a nuanced analysis of content authenticity and type. 
SynHate comprises over 134,000 samples drawn from MuTox and ADIMA datasets. It covers 37 languages from diverse language families, reflecting the international scope of online communication.
The audio samples, standardized to 16kHz, 16-bit PCM WAV format, vary in duration averaging 5 seconds for MuTox-derived samples and 20 seconds for ADIMA-derived ones and are accompanied by comprehensive metadata including language, class labels, source dataset, and transcripts. This carefully curated dataset not only advances the technical capabilities of hate speech detection but also represents a crucial step toward safeguarding digital media integrity in an era of sophisticated synthetic content.

\section{The Proposed SynHate Dataset} 

The SynHate dataset represents a significant advancement in multilingual audio hate speech spoofing detection by addressing the critical intersection of hate speech, synthetic audio, and diverse linguistic content. By integrating hate speech detection with synthetic audio identification, SynHate provides a unified framework to counter the growing threat of AI-generated hate speech across various linguistic and cultural contexts. Motivated by the increasing prevalence of online hate speech and the emerging risks posed by advanced text-to-speech (TTS) technologies, this dataset fills a critical gap where existing resources fall short. The proposed dataset is created by leveraging two established hate speech datasets: MuTox~\cite{costa2024mutox} and ADIMA~\cite{gupta2022adima}. The baseline code and the SynHate dataset is publicly available on \url{https://www.iab-rubric.org/resources} . Below, we describe the generation policies and procedures for the SynHate dataset.

\subsection{Dataset Generation}

SynHate is generated using two primary datasets: (i) \textbf{MuTox:} Selected for its extensive coverage of 30 languages, MuTox reflects the global nature of online communication. Its linguistic diversity includes major world languages such as English and Spanish, as well as less commonly studied languages like Hebrew, Bulgarian, and Swahili, enabling models that generalize across diverse cultural contexts. (ii) \textbf{ADIMA:} To address the unique challenges of low-resource languages, ADIMA is incorporated. This dataset contains speech samples from 10 Indic languages; however, due to compatibility issues with the speech-to-text model, Bhojpuri was excluded, leaving 9 effective languages. This focus is particularly valuable given the rapid growth of internet usage in South Asia. The complete pipeline for dataset generation is illustrated in Figure~\ref{fig:proposed_dataset_pipeline}.

\subsection{Database Generation Process}

The construction of SynHate involved a multi-stage process tailored to the unique characteristics of the source datasets:

\begin{enumerate}
    \item \textbf{Source Dataset Selection and Preprocessing:}
    The MuTox dataset, encompassing 30 languages, served as one of the primary sources for SynHate. Provided in TSV format, it contains hyperlinks to audio clips, timestamps, language information, and additional metadata. Audio clips inaccessible due to network errors or incorrect timestamps were excluded. The ADIMA dataset, focusing on 10 Indic languages (with Bhojpuri excluded), was accessed via a Google Drive link, providing audio clips along with class labels (abusive/non-abusive) and language information. ADIMA audio clips average around 20 seconds, longer than those in MuTox. Audio clips from MuTox and ADIMA were trimmed and padded to ensure fixed uniform durations: 5 seconds for MuTox and 20 seconds for ADIMA
    
    
    \item \textbf{Language Identification:}
    An optional Language Identification (LID) step using the MMS model, which supports over 4,017 languages, was incorporated to verify and, if necessary, correct the language tags provided in the source datasets.
    
    \item \textbf{Speech-to-Text Conversion:}
    While the MuTox dataset includes original text transcriptions for its audio clips, ADIMA required the use of the Speech-to-Text (STT) component of the MMS model~\cite{pratap2023mms} to generate transcripts. This conversion was crucial for enabling the subsequent synthetic generation.

    \item \textbf{Synthetic Speech Generation:}
    The text transcripts were processed using the Text-to-Speech (TTS) component of the MMS-1B (Massively Multilingual Speech) model. This state-of-the-art system generates high-quality synthetic speech across multiple languages. Quality control measures were applied to ensure that the generated fake audio maintained a high standard, thus creating a reliable parallel set of synthetic samples for each genuine audio clip.
\end{enumerate}

\begin{table*}[t!]
\centering
\caption{Statistics of the proposed Synhate dataset. The dataset is created using two source datasets, MuTox and ADIMA.}
\label{tab:dataset_statistics}
\begin{tabular}{|l|ccc|cc|}
\hline
\multirow{2}{*}{\textbf{Class}} & \multicolumn{3}{c|}{\textbf{MuTox Subset}}                                    & \multicolumn{2}{c|}{\textbf{ADIMA Subset}} \\ \cline{2-6} 
                       & \multicolumn{1}{c|}{\textbf{Train}} & \multicolumn{1}{c|}{\textbf{Val}} & \textbf{Test}  & \multicolumn{1}{c|}{Train} & \textbf{Test} \\ \hline
Real-Normal            & \multicolumn{1}{c|}{28895} & \multicolumn{1}{c|}{7913}       & 4425  & \multicolumn{1}{c|}{4358}  & 1934 \\ \hline
Real-Hate              & \multicolumn{1}{c|}{5018}  & \multicolumn{1}{c|}{861}        & 767   & \multicolumn{1}{c|}{3011}  & 1377 \\ \hline
Fake-Normal            & \multicolumn{1}{c|}{38891} & \multicolumn{1}{c|}{12928}      & 6105  & \multicolumn{1}{c|}{3642}  & 1594 \\ \hline
Fake-Hate              & \multicolumn{1}{c|}{6246}  & \multicolumn{1}{c|}{1488}       & 973   & \multicolumn{1}{c|}{3000}  & 1371 \\ \hline
Total                  & \multicolumn{1}{c|}{79050} & \multicolumn{1}{c|}{23190}      & 12270 & \multicolumn{1}{c|}{14011} & 6276 \\ \hline
\end{tabular}

\end{table*}

\subsection{Dataset Statistics}

The SynHate dataset combines the strengths of MuTox and ADIMA to offer a comprehensive resource for multilingual audio hate speech and spoofing detection. The dataset maintains the original train-validation-test splits to ensure consistency with previous research. Detailed statistics are provided in Table \ref{tab:dataset_statistics}. Key features include:

\begin{itemize}
    \item \textbf{Language Coverage:} 37 languages from diverse families (Indo-European, Semitic, Uralic, Dravidian), with 30 languages from MuTox and 9 from ADIMA.
    \item \textbf{Class Categorization:} A four-class system, Real-normal (RN), Real-hate (RH), Fake-normal (FN), and Fake-hate (FH), enables detailed analysis of authentic and synthetic hate speech.
    \item \textbf{Sample Size:} Approximately 134,797 total samples (114,510 from MuTox, 20,287 from ADIMA).
    \item \textbf{Audio Format:} Standardized to 16kHz, 16-bit PCM WAV format.
    \item \textbf{Metadata:} Each sample includes language information, class labels, source dataset, and transcripts (both original and generated).
\end{itemize}

\section{Experimental Protocols}
This section describes the training and testing protocols implemented on the SynHate dataset, and details the baseline algorithms along with their implementation for benchmarking. Our experimental setup aims to address the following key research questions.

\begin{itemize}
    \item \textbf{RQ1: Multilingual Spoofed Hate Speech Detection} \\
    Can the SynHate dataset be effectively utilized to detect deepfake hate audio across multiple languages?
    
    \item \textbf{RQ2: Impact of Language Diversity on Detection} \\
    Does training on global languages enhance the detection of synthetic hate speech in Indic languages, and vice versa?
\end{itemize}

\subsection{Baselines and Implementation Details}
To establish a comprehensive baseline for evaluation, we benchmarked the SynHate dataset using five state-of-the-art self-supervised models that represent a diverse array of architectures and training strategies:

\begin{itemize}
    \item \textbf{Whisper-small (244M) and Whisper-medium (769M)}~\cite{radford2023robust}: These models employ weakly supervised pre-training on 680,000 hours of multilingual audio, providing robust performance across diverse linguistic inputs.
    \item \textbf{XLS-R (965M)}~\cite{babu2021xls}: Based on the wav2vec 2.0 architecture, XLS-R leverages self-supervised learning on 436,000 hours of multilingual audio, offering strong representation learning capabilities.
    \item \textbf{AST}~\cite{gong2021ast}: A convolution-free model that utilizes a self-attention mechanism and is fine-tuned on AudioSet, demonstrating effectiveness in audio classification tasks.
    \item \textbf{mHuBERT (95M)}~\cite{boito2024mhubert}: This model adopts a modified HuBERT architecture and benefits from self-supervised learning on 90,000 hours of multilingual audio.
\end{itemize}

All models were fine-tuned using their respective pre-trained weights for ten epochs. We employed the cross-entropy loss function and the Adam optimizer with a learning rate of 0.0001. This diverse set of models, encompassing encoder-decoder frameworks, self-attention mechanisms, and self-supervised learning techniques, offers a thorough baseline for evaluating the performance of multilingual spoofed hate speech detection using the SynHate dataset.


\begin{table}[t!]
\centering
\caption{Summarizing the accuracies of baseline models on the Synhate dataset.}
\label{tab:baseline_preformnace}
\resizebox{\columnwidth}{!}{%
\begin{tabular}{|l|cc|cc|}
\hline
\multirow{2}{*}{\textbf{Model}} & \multicolumn{2}{c|}{\textbf{MuTox Subset}}        & \multicolumn{2}{c|}{\textbf{ADIMA Subset}}        \\ \cline{2-5} 
                       & \multicolumn{1}{c|}{\textbf{Accuracy}} & \textbf{F1 Score} & \multicolumn{1}{c|}{\textbf{Accuracy}} & \textbf{F1 Score} \\ \hline
AST                    & \multicolumn{1}{c|}{82.2}     & 0.824    & \multicolumn{1}{c|}{77.2}     & 0.773    \\ \hline
XLS-R                  & \multicolumn{1}{c|}{84.6}     & 0.843    & \multicolumn{1}{c|}{77.5}     & 0.777    \\ \hline
mHuBERT                & \multicolumn{1}{c|}{84.6}     & 0.855    & \multicolumn{1}{c|}{84.7}     & 0.847    \\ \hline
Whisper-small          & \multicolumn{1}{c|}{85.4}     & 0.865    & \multicolumn{1}{c|}{85.2}     & 0.849    \\ \hline
Whisper-medium         & \multicolumn{1}{c|}{83.5}     & 0.855    & \multicolumn{1}{c|}{83.1}     & 0.831    \\ \hline
\end{tabular}%
}
\end{table}
\begin{table}[t!]
\centering
\caption{Summarizing the language-wise accuracies of hate speech detection on the ADIMA subset.}
\label{tab:adima_language}
\resizebox{\columnwidth}{!}{%
\begin{tabular}{|l|c|c|c|c|c|}
\hline
\multirow{2}{*}{\textbf{Languages}}        & \multirow{2}{*}{\textbf{AST}}  & \multirow{2}{*}{\textbf{m-HuBERT}} & \multirow{2}{*}{\textbf{XLS-R}} & \textbf{Whisper} & \textbf{Whisper} \\ 
         &   &  &  & \textbf{-small} & \textbf{-medium} \\ \hline
Bengali   & 0.78 & 0.80     & 0.77  & 0.83          & 0.80           \\ \hline
Gujarati  & 0.77 & 0.82     & 0.75  & 0.85          & 0.83           \\ \hline
Haryanvi  & 0.80 & 0.87     & 0.79  & 0.88          & 0.86           \\ \hline
Hindi     & 0.75 & 0.85     & 0.76  & 0.86          & 0.87           \\ \hline
Kannada   & 0.73 & 0.83     & 0.76  & 0.83          & 0.81           \\ \hline
Malayalam & 0.79 & 0.84     & 0.79  & 0.85          & 0.84           \\ \hline
Odia      & 0.78 & 0.82     & 0.81  & 0.84          & 0.80           \\ \hline
Punjabi   & 0.80 & 0.87     & 0.80  & 0.86          & 0.84           \\ \hline
Tamil     & 0.76 & 0.83     & 0.75  & 0.85          & 0.82           \\ \hline
\end{tabular}%
}
\end{table}

\section{Results and Analysis}

We evaluate the performance of several baseline models on the SynHate dataset for a four-class classification task that distinguishes between real-normal, real-hate, fake-normal, and fake-hate speech. The models were trained on the designated training set and assessed on the test set, with detailed performance metrics presented in Table~\ref{tab:baseline_preformnace}. Our analysis provides key insights into model performance across diverse languages and datasets, highlighting the challenges inherent in multilingual and cross-dataset hate speech detection.

\subsection{RQ1: Multilingual Spoofed Hate Speech Detection}

\subsubsection{Results on the MuTox Subset}
On the MuTox subset, which encompasses a broad range of global languages, the Whisper-small model achieved the best performance, with a test accuracy of 85.4\% and an F1-score of 0.865. In particular, Whisper-small outperformed its larger counterpart (Whisper-medium, which achieved 83.5\% accuracy), suggesting that a larger model size does not always correlate with improved detection of synthetic hate speech. Both XLS-R and mHuBERT achieved comparable accuracies (84.6\%), although mHuBERT attained a slightly higher F1-score of 0.855, indicating its potential advantage in handling nuanced detection of hate speech.

Analysis across language families reveals distinct performance patterns. For instance, Mandarin Chinese achieved a perfect accuracy (100\%) across all models, setting a benchmark for optimal performance. In contrast, while Slavic languages like Czech and Slovak consistently achieved above 93\% accuracy, Germanic languages exhibited variability with German models struggling (64--77\% accuracy) and Dutch performing moderately (84--89\% accuracy). Among Romance languages, Catalan performed exceptionally (94--97\% accuracy), whereas Spanish and Italian underperformed (71--75\% accuracy), potentially due to dialectal differences or variations in training data quality. Semitic languages (Hebrew and Arabic) and other Asian languages (e.g., Vietnamese and Indonesian) demonstrated robust performance, generally exceeding 85\% accuracy. Detailed language-specific results are provided in Table~\ref{tab:mutox_languages}.

\subsubsection{Results on the ADIMA Subset}
For the ADIMA subset, focused on Indic languages, the performance trends differ slightly. Whisper-small again leads with a test accuracy of 85.2\% and an F1-score of 0.849, followed closely by mHuBERT (accuracy and F1-score of 84.7\%). Whisper-medium records third-best performance (both metrics at 83.1\%), while XLS-R and AST lag behind with accuracies around 77.5\% and 77.2\%, respectively.
 
Within the ADIMA subset, Haryanvi and Punjabi consistently register the highest performance among Indic languages, with accuracies in the range of 80--88\%. Dravidian languages exhibit varied outcomes: Malayalam shows consistent performance (79--85\% accuracy), while Tamil and Kannada vary between 73\% and 85\%. Indo-Aryan languages also display diversity, with Bengali achieving 78--83\% accuracy and Hindi ranging from 75\% to 87\%. Gujarati and Odia maintain stable performance, with accuracies around 77--85\% and 78--84\%, respectively. Detailed results by language are shown in Table~\ref{tab:adima_language}.

\subsection{RQ2: Impact of Languages on Detection}
\begin{table}[t!]
\centering
\caption{Summarizing the language-wise accuracies of hate speech detection on the MuTox subset.}
\label{tab:mutox_languages}
\resizebox{\columnwidth}{!}{%
\begin{tabular}{|l|c|c|c|c|c|}
\hline
\multirow{2}{*}{\textbf{Languages}}        & \multirow{2}{*}{\textbf{AST}}  & \multirow{2}{*}{\textbf{m-HuBERT}} & \multirow{2}{*}{XLS-R} & \textbf{Whisper} & \textbf{Whisper} \\ 
         &   &  &  &\textbf{ -small} &\textbf{ -medium} \\ \hline
Italian           & 0.75 & 0.75     & 0.69  & 0.75          & 0.81           \\ \hline
Russian           & 0.91 & 0.93     & 0.94  & 0.94          & 0.94           \\ \hline
Hungarian         & 0.84 & 0.87     & 0.86  & 0.89          & 0.87           \\ \hline
Bulgarian         & 0.83 & 0.84     & 0.83  & 0.84          & 0.82           \\ \hline
Hindi             & 0.84 & 0.85     & 0.88  & 0.88          & 0.88           \\ \hline
Czech             & 0.96 & 0.96     & 0.96  & 0.96          & 0.96           \\ \hline
vietnamese        & 0.85 & 0.82     & 0.88  & 0.88          & 0.88           \\ \hline
French            & 0.90 & 0.88     & 0.92  & 0.92          & 0.91           \\ \hline
Swahili           & 0.91 & 0.93     & 0.93  & 0.93          & 0.94           \\ \hline
Catalan           & 0.94 & 0.97     & 0.96  & 0.97          & 0.96           \\ \hline
Danish            & 0.83 & 0.86     & 0.84  & 0.85          & 0.84           \\ \hline
Estonian          & 0.84 & 0.88     & 0.87  & 0.87          & 0.88           \\ \hline
English           & 0.79 & 0.82     & 0.83  & 0.84          & 0.79           \\ \hline
Spanish           & 0.71 & 0.73     & 0.72  & 0.73          & 0.71           \\ \hline
Chinese (Mandarin) & 1.00 & 1.00     & 1.00  & 1.00          & 1.00           \\ \hline
Western Persian   & 0.93 & 0.94     & 0.93  & 0.95          & 0.94           \\ \hline
Polish            & 0.88 & 0.93     & 0.91  & 0.93          & 0.91           \\ \hline
German            & 0.64 & 0.75     & 0.74  & 0.77          & 0.75           \\ \hline
Urdu              & 0.83 & 0.79     & 0.86  & 0.84          & 0.85           \\ \hline
Arabic            & 0.89 & 0.89     & 0.88  & 0.89          & 0.90           \\ \hline
Hebrew            & 0.94 & 0.96     & 0.95  & 0.96          & 0.95           \\ \hline
Finnish           & 0.85 & 0.89     & 0.89  & 0.91          & 0.89           \\ \hline
Bengali           & 0.94 & 0.97     & 0.97  & 0.98          & 0.97           \\ \hline
Slovak            & 0.93 & 0.96     & 0.95  & 0.96          & 0.96           \\ \hline
Portuguese        & 0.77 & 0.79     & 0.81  & 0.83          & 0.81           \\ \hline
Greek             & 0.88 & 0.90     & 0.90  & 0.91          & 0.89           \\ \hline
Turkish           & 0.93 & 0.91     & 0.93  & 0.91          & 0.93           \\ \hline
Indonesian        & 0.88 & 0.90     & 0.92  & 0.92          & 0.91           \\ \hline
Tagalog           & 0.93 & 0.93     & 0.93  & 0.94          & 0.94           \\ \hline
Dutch             & 0.84 & 0.86     & 0.89  & 0.85          & 0.88           \\ \hline
\end{tabular}%
}
\end{table}
\subsubsection{Cross-Subset Evaluation}
We also conducted cross-dataset evaluations to understand how models trained on one subset perform on the other. Notably, the Whisper-small model trained on the MuTox subset achieved a cross-dataset test accuracy of 50.2\% and an F1-score of 41.9\% when tested on ADIMA. Conversely, among models trained on ADIMA, mHuBERT performed best when evaluated on MuTox, with an accuracy of 51.6\% and an F1-score of 61.6\%. In both scenarios, none of the models exceeded a 52\% accuracy threshold in cross-dataset settings, reflecting substantial differences in hate speech characteristics between the two subsets.

\subsubsection{Analysis of Cross-Dataset Performance}
The observed performance gap can be attributed to differences in the intensity and nature of hate speech between ADIMA and MuTox. ADIMA generally contains more intense expressions of hate speech, while MuTox represents a broader spectrum of linguistic nuances. Moreover, the models face challenges in generalizing across diverse cultural and linguistic contexts. For example, models trained on MuTox struggled to accurately classify both real and synthetic hate speech when tested on ADIMA. In particular, Haryanvi and Bengali exhibited the lowest cross-dataset performance, whereas Malayalam recorded the best results among ADIMA languages. Similarly, when models trained on ADIMA were tested on MuTox, languages such as Urdu, Hindi, and Russian showed relatively stronger performance, while Slovak and Czech consistently underperformed. Additionally, English and Spanish, which are prominent in MuTox, recorded accuracies below 50\% in cross-dataset evaluations, significantly affecting overall performance.

Overall, these results highlight the challenges in developing robust, cross-lingual hate speech detection systems and illustrate the value of the SynHate dataset in advancing research in this domain. The findings not only demonstrate the strengths of various baseline models in multilingual detection but also reveal limitations in current methodologies, paving the way for future research to bridge these gaps.

\begin{table}[t!]
\centering
\caption{Performance of baseline models when evaluated in cross-corpus settings.}
\label{tab:cross_val}
\resizebox{\columnwidth}{!}{%
\begin{tabular}{|l|cc|cc|}
\hline
\multirow{2}{*}{\textbf{Model}} & \multicolumn{2}{c|}{\begin{tabular}[c]{@{}c@{}}\textbf{Train on MutoX} \& \\ \textbf{Evaluation on ADIMA}\end{tabular}} & \multicolumn{2}{c|}{\begin{tabular}[c]{@{}c@{}}\textbf{Train on ADIMA }\& \\ \textbf{Evaluation on MutoX}\end{tabular}} \\ \cline{2-5} 
                       & \multicolumn{1}{c|}{\textbf{Accuracy}}                                & \textbf{F-1}                               & \multicolumn{1}{c|}{\textbf{Accuracy}}                                & \textbf{F-1                               }\\ \hline
AST                    & \multicolumn{1}{c|}{24.70}                                   & 0.28                                   & \multicolumn{1}{c|}{31.50}                                   & 0.37                                   \\ \hline
XLS-R                  & \multicolumn{1}{c|}{40.10}                                   & 0.36                                   & \multicolumn{1}{c|}{20.10}                                   & 0.28                                   \\ \hline
mHuBERT                & \multicolumn{1}{c|}{35.70}                                   & 0.33                                   & \multicolumn{1}{c|}{51.60}                                   & 0.62                                   \\ \hline
Whisper-small          & \multicolumn{1}{c|}{50.10}                                   & 0.42                                   & \multicolumn{1}{c|}{48.10}                                   & 0.50                                   \\ \hline
Whisper-medium         & \multicolumn{1}{c|}{43.00}                                   & 0.39                                   & \multicolumn{1}{c|}{46.00}                                   & 0.50                                   \\ \hline
\end{tabular}%
}
\end{table}


\section{Conclusion and Future Directions}

The SynHate dataset introduced in this paper represents a significant advancement in multilingual audio hate speech spoofing detection, addressing critical gaps in existing resources. By combining hate speech detection with audio spoofing detection, SynHate offers a novel four-class categorization system: real-normal, real-hate, fake-normal, and fake-hate across more than 134,000 samples and 37 languages. The integration of the MuTox and ADIMA datasets provides a rich resource for examining hate speech patterns in both global and Indic languages. Our baseline evaluations using five state-of-the-art models reveal varying performance across languages, with the Whisper-small model achieving the highest overall accuracy, and highlight the complexities inherent in cross-linguistic hate speech detection as well as the challenges posed by cross-dataset generalization.

This work opens several exciting avenues for future research. Expanding the dataset to include additional languages, dialects, and multimodal data sources, such as text and visual content, could further enrich the analysis and detection capabilities. Exploring self-supervised and ensemble methods, as well as advanced domain adaptation, can further improve model robustness and generalization across languages and datasets. By providing this comprehensive dataset along with baseline evaluations, our objective is to catalyze further research in this critical field, paving the way for the development of more sophisticated, culturally aware, and robust detection systems that contribute to safer and more inclusive online environments across diverse global communities.

\section{Acknowledgement}
\vspace{-2pt}

This research is supported by a grant from IndiaAI and Meta via the Srijan: Centre of Excellence for Generative AI. Pipariya was supported by the ACM IKDD Uplink Internship.

\bibliographystyle{IEEEtran}
\bibliography{refs,strings}

\end{document}